\documentclass[preprintnumbers,article,amsmath,amssymb,floatfix,10pt,prd,twocolumn,
superscriptaddress,nofootinbib]{revtex4-2}
\usepackage{bm}
\usepackage{amsfonts}
\usepackage{latexsym}
\usepackage[latin1]{inputenc}
\usepackage{graphicx}
\usepackage{amsmath}
\usepackage{palatino}
\usepackage{mathpazo}
\usepackage{textcomp}
\usepackage{float}
\usepackage{booktabs}
\usepackage{dcolumn}
\usepackage{ragged2e}
\usepackage{hyperref}
\hypersetup{colorlinks,citecolor=blue}
\hypersetup{colorlinks=true,linkcolor=blue,filecolor=magenta,    urlcolor=blue}
\usepackage{amsmath}
\usepackage{subfigure}
\usepackage[]{natbib}
\usepackage{xcolor}
\usepackage{orcidlink}
\usepackage{epsfig}
\usepackage{caption}
\usepackage{commath}
\def\jnl@style{\it}
\def\aaref@jnl#1{{\jnl@style#1}}

\def\aaref@jnl#1{{\jnl@style#1}}

\def\aj{\aaref@jnl{AJ}}                   
\def\apj{\aaref@jnl{ApJ}}                 
\def\apjl{\aaref@jnl{ApJ}}                
\def\apjs{\aaref@jnl{ApJS}}               
\def\apss{\aaref@jnl{Ap\&SS}}             
\def\aap{\aaref@jnl{A\&A}}                
\def\aapr{\aaref@jnl{A\&A~Rev.}}          
\def\aaps{\aaref@jnl{A\&AS}}              
\def\mnras{\aaref@jnl{Mon.~Not.~Roy.~Astron.~Soc.}}             
\def\prd{\aaref@jnl{Phys.~Rev.~D}}        
\def\prc{\aaref@jnl{Phys.~Rev.~C}}  
\def\prl{\aaref@jnl{Phys.~Rev.~Lett.}}    
\def\qjras{\aaref@jnl{QJRAS}}             
\def\skytel{\aaref@jnl{S\&T}}             
\def\ssr{\aaref@jnl{Space~Sci.~Rev.}}     
\def\zap{\aaref@jnl{ZAp}}                 
\def\nat{\aaref@jnl{Nature}}              
\def\aplett{\aaref@jnl{Astrophys.~Lett.}} 
\def\apspr{\aaref@jnl{Astrophys.~Space~Phys.~Res.}} 
\def\physrep{\aaref@jnl{Phys.~Rep.}}      
\def\physscr{\aaref@jnl{Phys.~Scr}}       
\def\commat{\aaref@jnl{Comm.~Math.~Phys.}}              
\def\science{\aaref@jnl{Science}}               
\def\cqg{\aaref@jnl{Classical Quant.~Grav.}}            
\def\jpcs{\aaref@jnl{JPCS}}                                     
\def\ijmpd{\aaref@jnl{Int.~J.~Mod.~Phys.~D}}                    
\def\grg{\aaref@jnl{Gen.~Relat.~Gravit.}}               
\def\rpp{\aaref@jnl{Rep.~Prog.~Phys.}}          
\def\npa{\aaref@jnl{Nucl.~Phys.~A}}        
\def\lrr{\aaref@jnl{Living Rev.~Rel.}}                   
\def\jcap{\aaref@jnl{J.~Cosmology Astropart.~Phys.}}    
\def\rmp{\aaref@jnl{Rev.~Mod.~Phys.}}   
\def\epjc{\aaref@jnl{Eur.~Phys.~J.~C}} 
\def\plb{\aaref@jnl{~Phy.~Lett.~B}} 
\def\mpla{\aaref@jnl{Mod.~Phy.~Lett.~A}} 
\def\arxiv{\aaref@jnl{arxiv.org}}

\allowdisplaybreaks[1]

\begin{document}
\color{black}       
\title{Revisiting kink-like parametrization and constraints using OHD/Pantheon+/BAO samples}

\author{Simran Arora\orcidlink{0000-0003-0326-8945}}
\email{dawrasimran27@gmail.com}
\affiliation{Department of Mathematics, Birla Institute of Technology and
Science-Pilani,\\ Hyderabad Campus, Hyderabad-500078, India.}

\author{P.K. Sahoo\orcidlink{0000-0003-2130-8832}}
\email{pksahoo@hyderabad.bits-pilani.ac.in}
\affiliation{Department of Mathematics, Birla Institute of Technology and
Science-Pilani,\\ Hyderabad Campus, Hyderabad-500078, India.}
%

\begin{abstract}
We reexamine the kink-like parameterization of the deceleration parameter to derive constraints on the transition redshift from cosmic deceleration to acceleration. This is achieved using observational Hubble data, Type Ia Supernovae Pantheon+ samples and Baryon acoustic oscillations. In this parametrization, the value of the initial $q$ parameter is $q_{i}$, the final value is $q_f$, the present value is denoted by $q_{0}$, and the transition duration is given by $\alpha$. We perform our calculations using the Monte Carlo Markov Chain method, utilizing the emcee package. Under the assumption of a flat geometry, we constrain the range of possible values for three scenarios: when $q_{f}$ is unrestricted, when $q_{f}$ is equal to $-1$, and when $\alpha$ is $1/3$. This is done assuming that $q_{i}=1/2$. Here, we achieve that the $SN$ data fixes the free parameters tightly as in the flat $\Lambda$CDM for unrestricted $q_{f}$. In addition, if we fix $q_{f}=-1$, the model behaves well as the $\Lambda$CDM for the combined dataset. We also acquire the current value of the deceleration parameter, which is consistent with the latest results that assume the $\Lambda$CDM model. Furthermore, we observe a deviation from the standard $\Lambda$CDM model in the current model based on the evolution of $j(z)$, and it is evident that the universe transitions from deceleration to acceleration and will eventually reach the $\Lambda$CDM model in the near future. \\

\end{abstract}

\maketitle

\tableofcontents

\section{Introduction}
The standard cosmological model is challenged by significant physical inconsistency caused by the presence of mysterious dark energy and dark matter, the characteristics of which are currently the subject of extensive theoretical discussion. The most significant conundrum in the standard cosmological model is the possibility that 75\%  of the universe content consists of dark energy, which is responsible for the accelerated expansion of the universe. In contrast, dark matter accounts for approximately 20\% of the universe content and is an unknown component that is crucial for the formation of structures. Gaining an in-depth understanding of both these components poses a major challenge within the standard cosmological model, also known as the $\Lambda$CDM paradigm \cite{Steinhardt/2011}.\\
While $\Lambda$CDM cosmology aligns with most observational data, it encounters significant challenges, notably fine-tuning \cite{Weinberg/1989}. According to observations, the magnitude of $\Lambda$ is very low, sufficient to drive this current accelerating phase. Whereas the theoretically predicted value of $\Lambda$, estimated from the quantum theory of fields, is enormously higher, and consequently, there is a huge discrepancy of the order of $10^{120}$ between the observed and theoretical estimations of $\Lambda$. Secondly, $\Lambda$ and matter magnitudes are unexpectedly comparable, leading to a coincidence problem \cite{Steinhardt/1999}. Consequently, these challenges have stimulated the quest for alternative physical explanations for explaining the dynamics of the universe, resulting in plenty of alternative approaches \cite{Yoo/2012}. The present issues highlight the necessity for additional frameworks that include the $\Lambda$CDM model \cite{Valentino/2020,Benetti/2019}. Supernovae Ia (SNe Ia) observations \cite{Riess/1998,Perlmutter/1999} provided the first evidence that the universe is accelerating, which was further supported by data from cosmic microwave background (CMB) radiation \cite{Komatsu/2011,Planck/2014}, baryonic acoustic oscillations \cite{Eisenstein2005,Percival/2007}, and the observed Hubble measurements \cite{Farooq/2017,Yu_2018}. \\
A further approach to understanding cosmic acceleration is to examine kinematic variables such as the Hubble parameter $H$, the deceleration parameter $q$, or the jerk parameter $j$, which are derived from the derivatives of the scale factor. The kinematic technique is helpful since it does not require any model-specific assumptions, such as the composition of the universe. A metric theory of gravity describes it, and it is assumed that the universe is homogeneous and isotropic at cosmological scales. Numerous attempts in the literature have been made to constrain the current values of $H$, $q$, and $j$ by parametrizing $q$ or $j$ \cite{Mukherjee,Nair,Mukherjee/2017}. Already, there are some relevant investigations in this direction. Several works in the literature estimated cosmological parameters independently of energy content, and some authors used parametrization in these estimates \cite{Asvesta/2002,Gaurav/2022,Pacif/2021}. Riess et al. \cite{Riess/2004} used a simple linear redshift parameterization of $q$, i.e., $q(z) = q_0 + q_{1}\,z$, to measure a transition from an early decelerating to a current accelerating phase. However, at high redshift, this model is unreliable. In a recent study, Xu et al. \cite{Xu/2009} considered the linear first-order expansion of $q$, i.e., $q(z) = q_0 + \frac{q_1\,z}{1+z}$, constant jerk parameter, and third-order growth of luminosity distance. Using a $q(z)$ parametrization, in contrast to a parametrization for the equation of state, offers the advantage of broader generality. In this scenario, we are not limited to the principles of general relativity, and the assumptions made about dark matter and energy are kept to a minimum. We simply need to adopt a metric theory of gravity.\\
We aim to find a parametrization that accurately represents several cosmological models across a broad range of redshift values. In the same vein, the major purpose of this study is to investigate several simple kinematic models for cosmic expansion based on a kink-like parameterization for $q(z)$ \cite{Ishida/2008}. There are some special cases described by our parametrization, such as a flat standard cosmological model with a constant dark energy equation of state ($\omega$CDM), $\Lambda$CDM, the flat DGP braneworld model, the flat quartessence Chaplygin model, modified Polytropic Cardassian model \cite{Dvali}. The kink-like parametrization has been investigated using SNeIa: the Gold182 sample processed with the light curve fitter MLCS2k2 and the SNLS survey, analyzed with SALT. There is tension between the results obtained using the GOLD182 dataset and those obtained with the SNLS dataset. It is seen that the SN observations in their work cannot impose strong constraints on the value of $z_{t}$. They remarked that even in the region of interest $(z_{t} \leq 1)$ the difference between the outcomes of the two SNeIa datasets, i.e. SNLS and Gold182, exists and can be related to possible inhomogeneities present in the Gold182 sample. The authors have assumed the old Hubble dataset, which comprises 28 points, and considered the values of $H_{0}$ as uniform priors to constrain the other parameters.  \\
Our primary focus lies in the transition from cosmic deceleration to acceleration. We are interested in the redshift of this transition, its duration, and the current value of the deceleration parameter. To constrain the parameters of our model, we utilize Observational Hubble data (OHD), Supernovae Pantheon+ samples, and Baryon Acoustic Oscillations (BAO). We additionally constrain the value of $H_0$ by utilizing datasets rather than assuming uniform or Gaussian priors, as done in previous studies. However, our work is more general and different from similar works in different ways. Firstly, we do not
assume the $H_{0}$ as a prior but rather allow our model to behave more generally through observational data. Secondly, in this work, we go one step further by studying the evolution of the equation of state and jerk parameters for a general deceleration parameter. Lastly, we employ the latest Pantheon+ samples here.\\
The framework of this study is as follows: Section \ref{section 2} provides an overview of the background cosmology and introduces our kink-like parametrization along with its associated features. Section \ref{section 3} concisely explains the observational data used in our analysis. This section presents constraints on the model parameters determined by the data sets. It also addresses the potential model selection using the Akaike and Bayesian criteria. The cosmological model's kinematic parameters, including the deceleration parameter, equation of state parameter, and jerk parameter, are discussed in Section \ref{section 4}. Section \ref{section 5} contains a concise overview and remarks on the obtained outcomes.

\section{The $q(z)$ parametrization}
\label{section 2}
A spatially homogeneous and isotropic universe is described by the Friedmann-Lema\^{i}tre-Robertson-Walker (FLRW) metric
\begin{equation}
    \label{metric}
    ds^{2}= -c^{2} dt^{2}+a^{2}(t)\left[ \frac{dr^{2}}{1-kr^{2}} + r^{2} d\theta^{2} + r^{2} sin^{2} \theta d\phi^{2} \right],
\end{equation}
where $a(t)$ represents the scale factor and $k=0,\pm 1$ depicts the curvature. In this study, our primary focus is on a flat cosmological model. The Hubble parameter is defined as $H=\frac{\dot{a}}{a}$, where $\dot{a}$ denotes derivative with respect to the cosmic time. It is possible to rewrite all cosmological parameters as functions of redshift $z$, using $\frac{a_{0}}{a}=\frac{1}{1+z}$ ($a(0)=a_{0}$). \\
A dimensionless measure of the cosmic acceleration is determined by the deceleration parameter, given by
\begin{equation}
\label{qdef}
    q = -\frac{\ddot{a}}{a H^{2}}.
\end{equation}
Cosmological measurements suggest that the universe is currently experiencing an era of accelerated expansion, i.e., $q<0$. Nevertheless, this acceleration must have commenced in the near past and is not an enduring characteristic of evolution. This transition from a decelerated to an accelerated phase of expansion is characterized by a change in the signature of $q$, which occurs at some specific $z_{t}$, known as the deceleration-acceleration transition redshift.\\
Here, we adopt a flat cosmological model whose deceleration parameter can be described by the given equation after the dominance of radiation \cite{Ishida/2008}.   
\begin{equation}
\label{q}
q(z)= q_{f} + \frac{\left(q_{i}-q_{f}\right)}{1-\frac{q_{i}}{q_{f}}\left(\frac{1+z_{t}}{1+z} \right)^{1/\alpha}}.
\end{equation}
The parameter $z_{t}$ represents the redshift at which the transition occurs from cosmic deceleration to acceleration, whereas $\alpha$ denotes the width of the transition and is related to the jerk at the transition by $\frac{1}{\alpha} = \left(\frac{1}{q_{i}} - \frac{1}{q_{f}}\right)j(z_{t})$.  \\
We can rewrite the Hubble parameter using $q$ as  
\begin{equation}
\label{h1}
    H = H_{0}\, exp\int_{0}^{z} \frac{1+q(y)}{1+y} dy,
\end{equation}
which further leads to 
\begin{equation}
\label{H}
    H(z) = H_{0} (1+z)^{(1+q_{i})} \left[\frac{q_{i} \left(\frac{1+z_{t}}{1+z}\right)^{1/\alpha}-q_{f}}{q_{i} \left(1+z_{t}  \right)^{1/\alpha}-q_{f}}  \right]^{\alpha (q_{i}-q_{f})}.
\end{equation}
The final value of the deceleration parameter $q_{f}=q|_{z=-1}$ can  be linked to the current value $q_{0}= q|_{z=0}$ by 
\begin{equation}
\label{qf}
    q_{f} = \frac{q_{i}\left(1+z_{t} \right)^{1/\alpha}}{1-\frac{q_{i}}{q_{0}}\left[ 1-(1+z_{t})^{1/\alpha} \right]}.
\end{equation}
As previously stated, our objective with the $q(z)$ parametrization is to describe the transition from a decelerated phase to an accelerated phase. Therefore, given that $q_{i}>0$ and $q_{f}<0$, it is straightforward to show that the parameter $\alpha$ is constrained to the range $0<\alpha<\frac{ln(1+z_{t})}{ln\left(1-\frac{q_{0}}{q_{i}} \right)}$.

In most cosmological scenarios, the formation of large-scale structures necessitates that the universe, during its early stages, undergoes a period dominated by matter, where $q=\frac{1}{2}$. Given that our parametrization of $q(z)$ is specifically intended to reflect the evolution of the universe starting from a phase dominated by matter and experiencing deceleration, we set $q_{i}=\frac{1}{2}$. It is important to note that other flat cosmological models explored in the literature can be seen as specific instances of kink-like parametrization, such as $\omega$CDM, $\Lambda$CDM, etc. It is seen that while our parametrization is true at high redshift, it is not applicable during the radiation-dominated era when $q=1$.\\
We aim to provide a sensible approach for determining constraints on the unknown parameters $(H_{0},q_{0},z_{t},\alpha)$.  

\section{Observational Data Analysis}
\label{section 3}
Within this section, we employ several combinations of datasets, such as the OHD and SN Pantheon+ data, to reconstruct the cosmic deceleration parameter $q$ as a function of the redshift $z$. In addition, we perform a reconstruction of $q$ using the combined $OHD$, $SN$, and $BAO$ datasets. To explore the parameter space, we will be using the MCMC methodology and Python package \texttt{emcee} \citep{emcee}.

\subsection{Observational Hubble Data}

Observational Hubble Data is a widely accepted and significant method for examining the expansion history of the universe beyond GR and $\Lambda$CDM. The OHD sample mainly derives from the differential age of galaxies method, often known as DAG \cite{Yu_2018}. The Hubble rate is usually obtained from the following formula
\begin{equation}
  H(z)=\frac{-1}{1+z}\frac{dz}{dt}.
\end{equation}
In this work, we mainly utilize the OHD points obtained from the Cosmic Chronometers (CC), which refer to the massive and passively evolving galaxies. By employing CC, one can get the rate of change $dz/dt$ by calculating $\Delta z/\Delta t$, where $\Delta z$ is the redshift separation in the galaxies sample. This value can be established with precision and accuracy through precise spectroscopy. However, obtaining $\Delta z$ is considerably more complex and requires conventional clocks. To accomplish this, we might employ old, massive, stellar populations that are passively evolving and exist across a broad spectrum of redshifts. To determine the priors and likelihood functions (which are necessary), we take into account the compilation consisting of 31 points. To constrain our model, we introduce the chi-square function
\begin{equation}
    \chi^2_{OHD}=\sum_{i=1}^{N_H}\bigg[\frac{H^\mathrm{th}_i(\Theta,z_i)-H^\mathrm{obs}_i(z_i)}{\sigma_{H(z_i)}}\bigg]^2,
\end{equation}
where $\Theta$ is a parameter space, $N_H$ denotes the number of data points, $z_{i}$ represent the redshift at which $H(z_i)$ has been measured, $H_{obs}$ and $H_{th}$ are the measured and predicted values of $H(z)$, respectively, and $\sigma_{H(z_{i})}$ is the standard deviation of the $i_{th}$ point.
The likelihood function that we will be using for MCMC sampling has its usual exponential form
\begin{equation}
    \mathcal{L}=\exp(-\chi^2/2).
\end{equation}

\subsection{Pantheon+ SNeIa Sample}
The Pantheon+ dataset comprises distance moduli calculated from 1701 light curves of 1550 SNeIa. These light curves were obtained from 18 different surveys, and they cover a redshift range of $0.001 \leq z \leq 2.2613$. It is worth noting that 77 out of the 1701 light curves are associated with Cepheid-containing galaxies \cite{Brout}. In the context of the Pantheon+ dataset, chi-square values are computed using the equation
\begin{equation}
    \chi^2_{SN}= \Delta \mu^{T} \left(C^{-1}_{stat+sys}\right) \Delta \mu,
\end{equation}
where $C^{-1}_{stat+sys}$ represents the covariance  matrix  of the Pantheon dataset, which combines both systematic and statistical
uncertainties, $\Delta \mu$ is the distance residual given by $\Delta \mu_{i}= \mu_{i}- \mu_{th}(z_{i})$. 
It is important to note that $\mu_{i}-m_{i}-M$, where $m_{i}$ is the apparent magnitude and $M$ is the fiducial magnitude of $i^{th}$ SNeIa. The distance moduli theoretically can be defined as 
\begin{equation}
    \mu^{th}=5\log_{10}d_L(z)+\mu_0,\quad \mu_0 = 5 \log_{10} \frac{H_0^{-1}}{\mathrm{Mpc}}+25,
\end{equation}
where
\begin{equation}
    d_L(z)=\frac{c(1+z)}{H_0}\bigg(H_0\int^z_0\frac{d\overline{z}}{H(\overline{z})}\bigg).
\end{equation}
The Pantheon+ dataset is an improvement over the previous Pantheon sample in that it resolves the ambiguity between the absolute magnitude $M$ and the Hubble constant $H_{0}$. This is accomplished by expressing distance moduli of SNeIa in the Cepheid host, defined as 
\begin{equation}
\Delta \tilde{\mu}=
    \begin{cases}
        \mu_{i} - \mu_{i}^{Ceph} & \text{if } i \in \text{Cepheid hosts}\\
        \mu_{i} - \mu_{th}^{z_{i}} & \text{Otherwise}
    \end{cases}
\end{equation}
where $\mu_{i}^{Ceph}$ represents Cepheid  host of the $i^{th}$ SNeIa which is provided by SH0ES. It is seen that $ \mu_{i} - \mu_{i}^{Ceph}$ is sensitive to $M$ and the Hubble constant $H_{0}$. Furthermore, in our analysis, we take $M=-19.523$, which has been determined from SH0ES Cepheid host distances \cite{Riess}.

\subsection{Baryon Acoustic Oscillations}
Finally, we constrain our model using Baryon Acoustic Oscillations (BAOs). BAOs appear early in the evolution of the universe. The so-called sound horizon $r_{s}$, which is visible at the photon decoupling epoch with redshift $z_{*}$, defines the characteristic scale of the BAO
\begin{equation}
    r_s=\frac{c}{\sqrt{3}}\int^{\frac{1}{1+z_*}}_0\frac{da}{a^2 H\sqrt{1+(3\Omega_{b0}/4\Omega_{\gamma0})a}}.
\end{equation}
In this case, the present baryon mass density is denoted by $\Omega_{b0}$, and $\Omega_{\gamma0}$ is the photon mass density at present. Here, $z_{*}$ is the redshift of photon decoupling, and we use $z_{*}=1091$. To obtain the BAO constraints, we use the ``acoustic scale" $l_{A} = \pi \frac{d_{A}(z_{*})}{r_{s}(z_{*})}$, where $d_A(z)=\int ^z_0\frac{dz'}{H(z')}$ is the angular diameter distance in the comoving coordinates. Here, $D_V(z_{BAO})$ is the dilation scale
\begin{equation}
    D_V(z)=\left[\frac{d_A(z)^2 cz}{H(z)}\right]^{1/3}.
\end{equation}
Finally, we obtain the BAO constraints $\left(\frac{d_{A}(z_{*})}{D_{V}(z_{BAO})}  \right)$ \cite{Planck2015}. This dataset was gathered from these references \cite{Blake,Eisenstein2005,Giostri2012}.

Consequently, to perform the MCMC sampling, we need to define the chi-square function for our BAO dataset
\begin{equation}
    \chi^2_{BAO}=X^T C^{-1}X,
\end{equation}
Where $X$ and $C^{-1}$ are of the form in \cite{Giostri2012}.\\

Additionally, we minimized $\chi^{2}_{OHD} +\chi^{2}_{SN} +\chi^{2}_{BAO}$ to carry out the joint analysis from the combined $OHD+SN+BAO$. The results are numerically derived from MCMC using Pantheon+, BAO, and joint (OHD+SN+BAO) datasets, which are presented in Table \ref{table1}. Furthermore, the $1-\sigma$ and $2-\sigma$ likelihood contours for the possible subsets of parameter space are presented for different cases in the next section. Here are the priors that we consider for free parameters
\begin{equation}
\begin{gathered}
H_0 \in [60,80], \quad q_{0}\in[-1,0],\\
z_{t} \in [0,1], \quad \alpha\in [0,0.8].\\
\end{gathered}
\end{equation}
These priors are extremely relaxed. Therefore, there is little to no bias. 

\subsection{Statistical evaluation}
To assess the effectiveness of our MCMC study, a statistical evaluation has to be conducted utilizing the Akaike Information Criterion (AIC) and Bayesian Information Criterion (BIC). The AIC can be expressed as follows $\mathrm{AIC} = \chi^2_{\mathrm{min}}+2d$ \cite{Akaike}, where $d$ is the number of free parameters in a considered model. To compare our results with the standard $\Lambda$CDM model, we use the AIC difference between our model and the standard cosmology $\Delta\mathrm{AIC}=|\mathrm{AIC}_{\Lambda\mathrm{CDM}}-\mathrm{AIC}_{\mathrm{Model}}|
$. Here, if $\Delta\mathrm{AIC}<2$, there is strong evidence in favor of our model, while for $4<\Delta\mathrm{AIC}\leq 7$, there is little evidence in favor of the model of our consideration. Finally, for the case with $\Delta \mathrm{AIC}>10$, there is practically no evidence in favor of our model \cite{Liddle}.\\
In addition, BIC is defined by $\mathrm{BIC} =\chi^2_{\mathrm{min}}+d\ln N$. Here, $N$ is the number of data points used for MCMC. For BIC, if $\Delta \mathrm{BIC}<2$, there is no strong evidence against a chosen model that deviates from $\Lambda$CDM, if $2\leq \Delta \mathrm{BIC}<6$, there is evidence against the model and finally for $\Delta\mathrm{BIC}>6$ there is strong evidence against our model. We therefore store the $\chi^2_{\mathrm{min}}$/AIC/BIC data for our model in the Table \ref{table1}.\\
We see that, for $q_{f}$ free case, $\Delta\rm AIC=0.83$ for $SN$, and $5.7$ for $OHD+SN+BAO$. In this case, the model strongly favours $SN$ data and is a little favourable for the joint samples. Whereas, in case of $q_{f}=-1$, $\Delta\rm AIC=2.56$ for $SN$ and $5.8$ for $OHD+SN+BAO$. In this case, the model favours all the combinations of datasets. According to BIC analysis, the present model for the case of $q_{f}=-1$ is consistent and favourable.  

\section{Results and Cosmological parameters} 
\label{section 4}

\subsection{When $q_{f}$ is free}

Let us examine the scenario when we set $q_{f}$ as a free parameter and establish $q_{i}=1/2$. It can assume any value. Instead of working with $q_{f}$, we use $q_{0}$. In general, three model parameters need to be constrained in this case, namely $(H_{0},q_{0},z_{t},\alpha)$. It is noteworthy that we explicitly use the restriction on $\alpha$, as mentioned before.

\begin{widetext}

\begin{figure}[]
   \begin{minipage}{0.48\textwidth}
     \centering
     \includegraphics[scale=0.45]{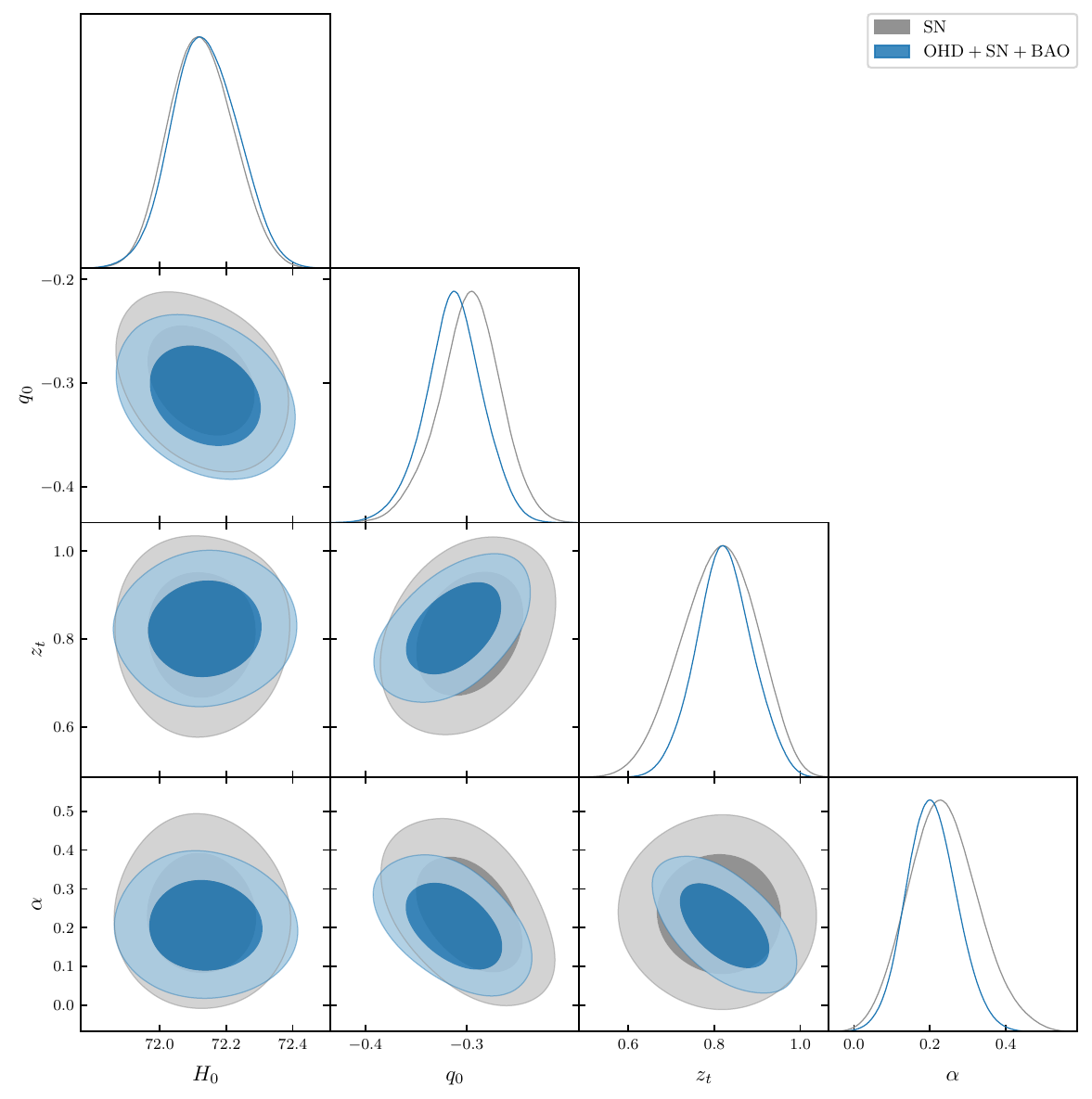}
     \caption{The marginalized posterior distributions in the parameter space $(H_{0}, q_{0}, z_{t},\alpha)$ for the case when $q_{f}$ is free.}\label{fig1}
   \end{minipage}\hfill
   \begin{minipage}{0.48\textwidth}
     \centering
     \includegraphics[scale=0.6]{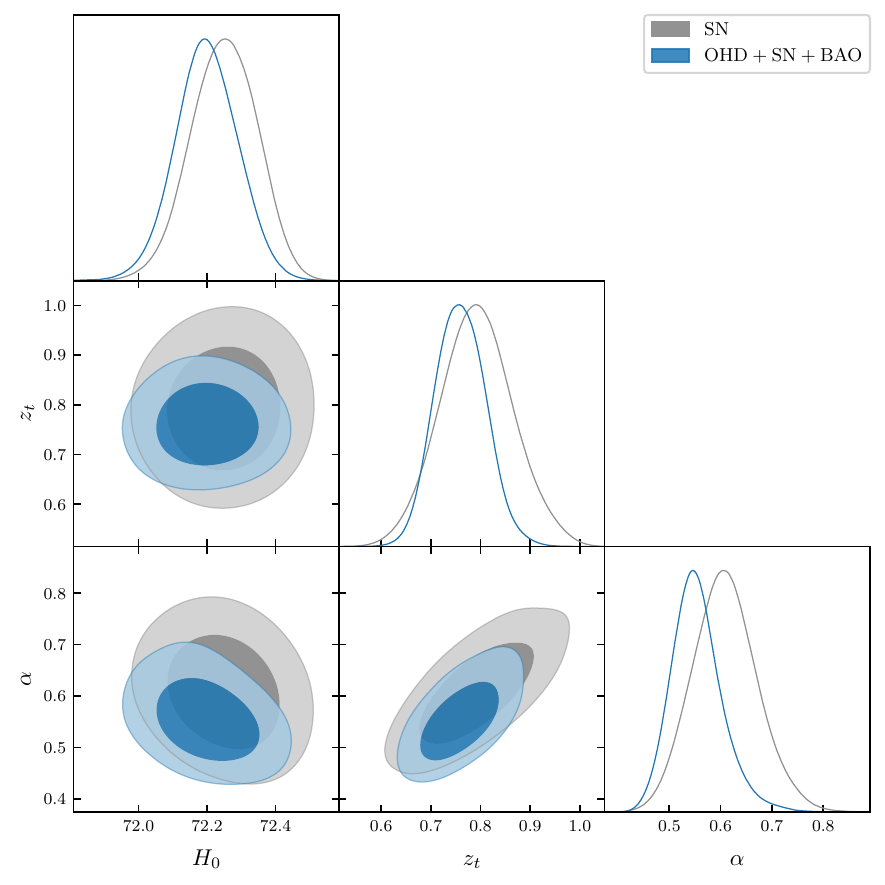}
     \caption{The marginalized posterior distributions in the parameter space $(H_{0}, z_{t},\alpha)$ for the case when $q_{f}=-1$.}\label{fig3}
   \end{minipage}
\end{figure}



\begin{figure}[]
\centering
\subfigure[]{\includegraphics[width=0.37\linewidth]{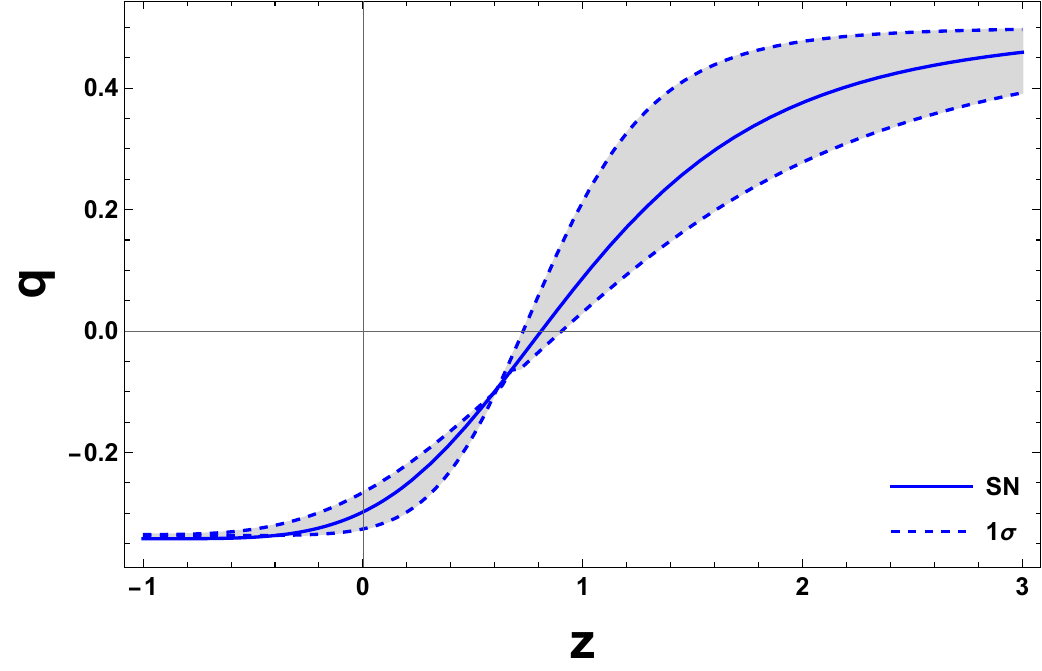}} \hspace{0.2in}
\subfigure[]{\includegraphics[width=0.37\linewidth]{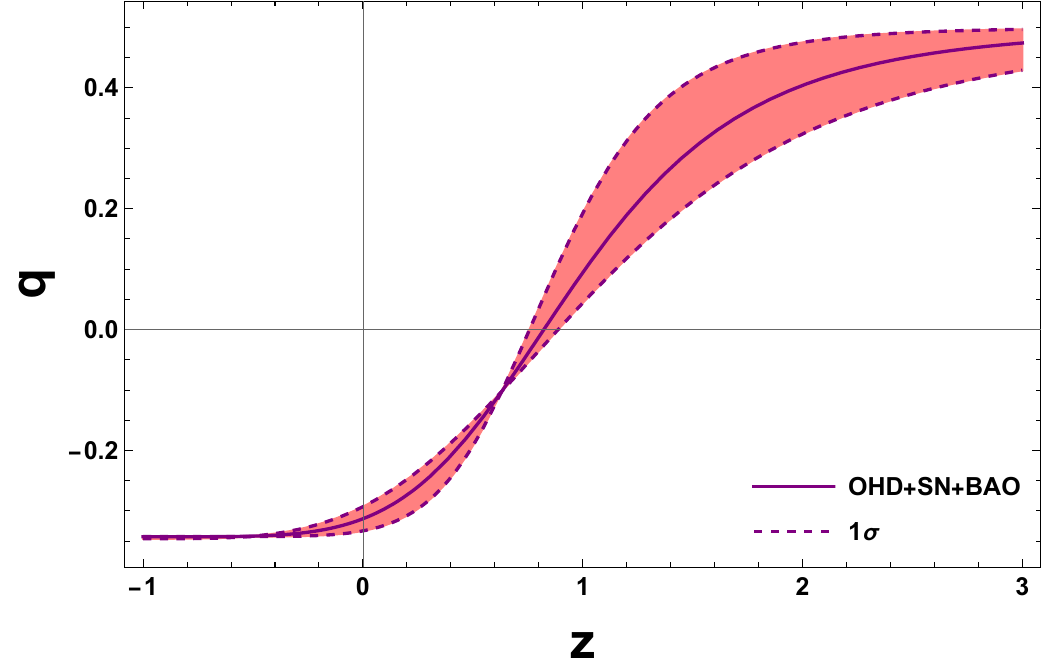}}\hspace{0.2in}
\subfigure[]{\includegraphics[width=0.37\linewidth]{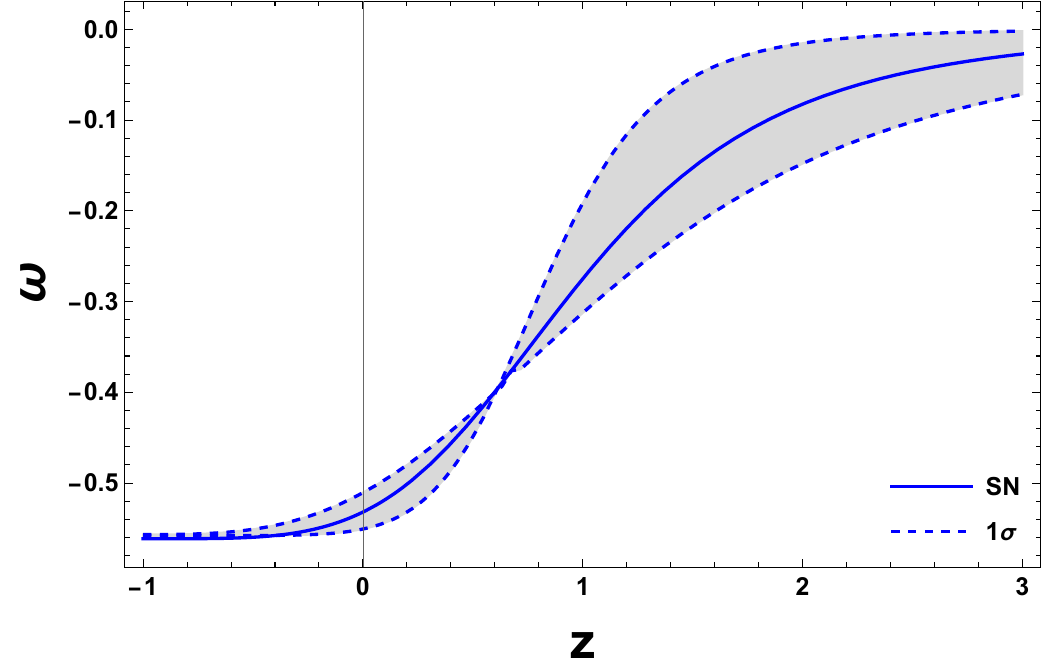}}\hspace{0.2in}
\subfigure[]{\includegraphics[width=0.37\linewidth]{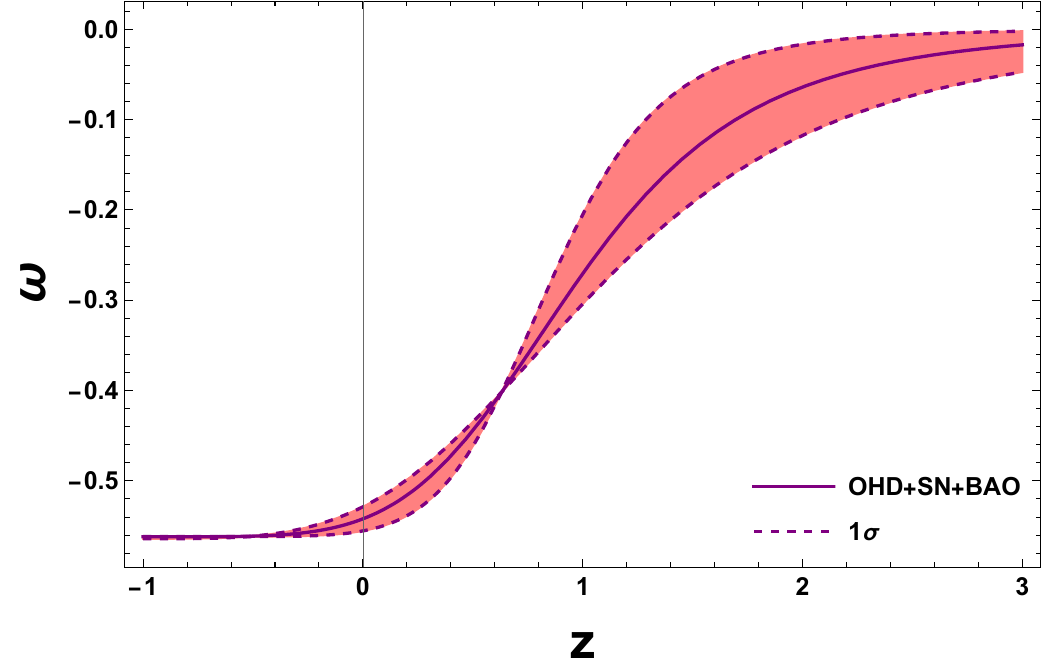}}
\caption{Behavior of the deceleration parameter and the equation of state parameter when $q_{f}$ is free.}
\label{fig2}
\end{figure}

\end{widetext}

In FIG. \ref{fig1}, we display 2D marginalized confidence level regions for parameter space $(H_{0}, q_{0}, z_{t},\alpha)$ using $SN$ and $OHD+SN+BAO$. Finally, we plot the deceleration parameter $q(z)$, the equation of state parameter $\omega(z)$ for all these considered data in FIG. \ref{fig2}. The behavior of $q(z)$ shows a signature flip, and $z_t$ indicates the epoch when the expansion of the universe went from a decelerating to an accelerating phase in the recent past. There is an uninspired phase of the universe when $z \rightarrow -1$ since $q(z)$ diverges more from the de-sitter phase. Another choice for $q_f$ could also be made to ensure the model's consistency in the distant future. Consequently, an additional analysis is required. The plot also displays the evolution of the equation of state parameter with respect to $z$. It reveals that the $\omega$ is positive at higher redshifts and remains greater than $-1$ (quintessence phase) until the present time for all the datasets. It does not cross a phantom divide. The present value of $\omega$ is $-0.532^{+0.021}_{-0.018}$ and $-0.54 \pm 0.13$, for $SN$ and $OHD+SN+BAO$, respectively. \\
However, the function $j(z) = q(2q+1) + (1+z) \frac{dq}{dz}$ demonstrates the deviation of $j$ from the flat $\Lambda$CDM model $(j = 1)$ in the best-fit model. The jerk parameter is often used to discriminate against various dark energy models. Observations indicate that, in this model, the current value of $j$ is far from $1$ for all the datasets ($j_{0}=0.05$ and $0.02$ for $SN$, $OHD+SN+BAO$, respectively). Therefore, based on the existing model, it is evident that the dynamic dark energy model considered is the most probable explanation for the current acceleration and needs attention. But in this case, the $j(z)$ is not well constrained as it may be connected to the appearance of abrupt future singularities \cite{Dabrowski, Pan/2018}.

\subsection{When $q_{f}$ is $-1$} 
\begin{widetext}

\begin{figure}[H]
\centering
\subfigure[]{\includegraphics[width=0.3\linewidth]{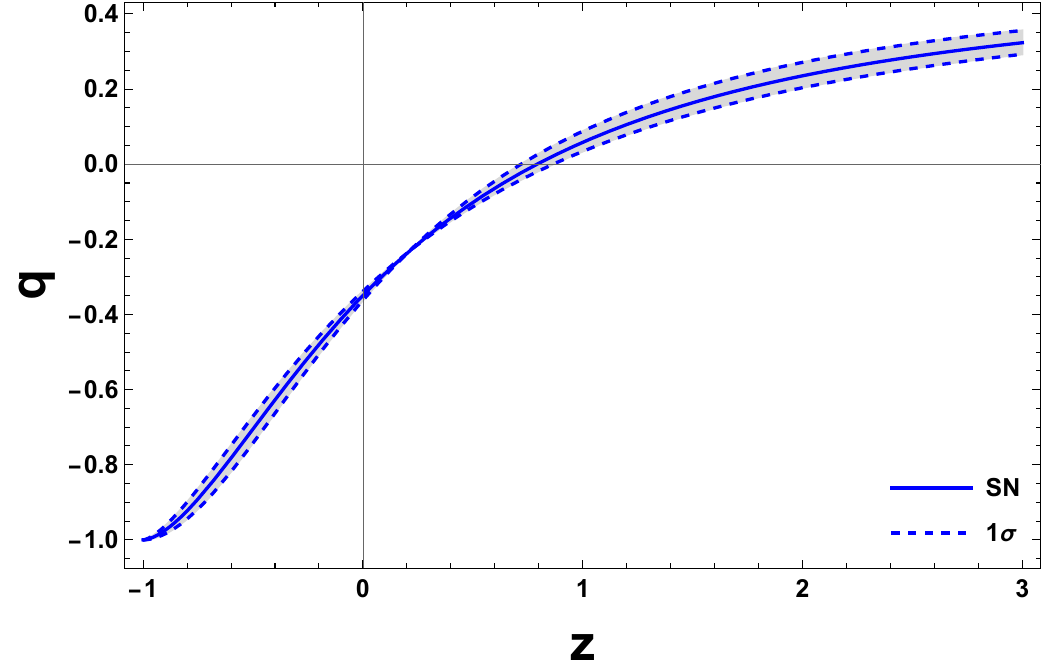}} \hspace{0.1in}
\subfigure[]{\includegraphics[width=0.3\linewidth]{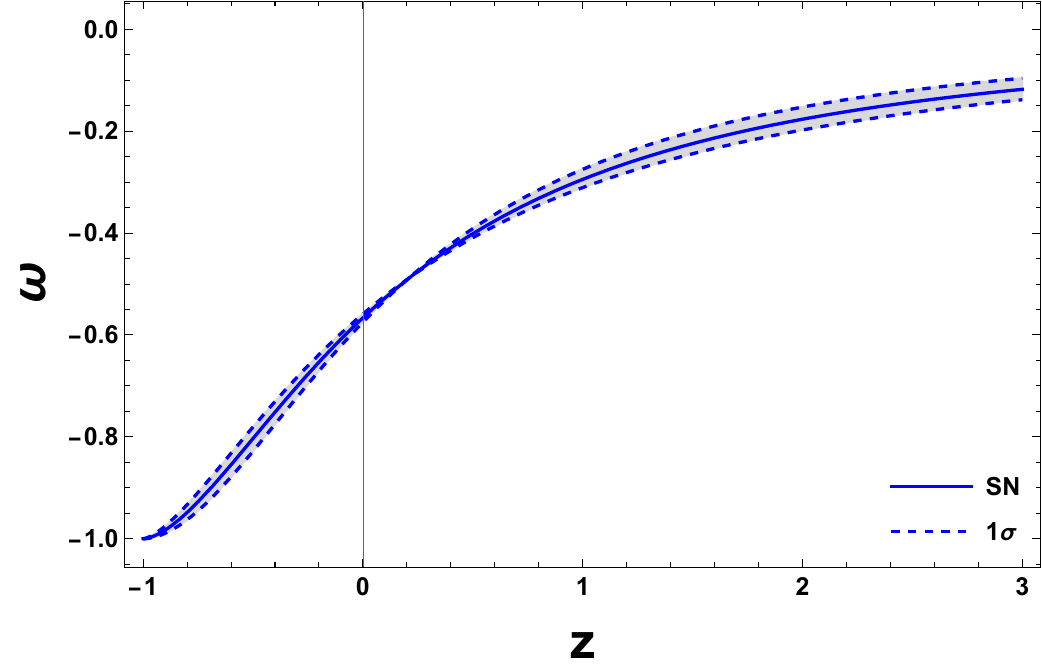}} \hspace{0.1in}
\subfigure[]{\includegraphics[width=0.3\linewidth]{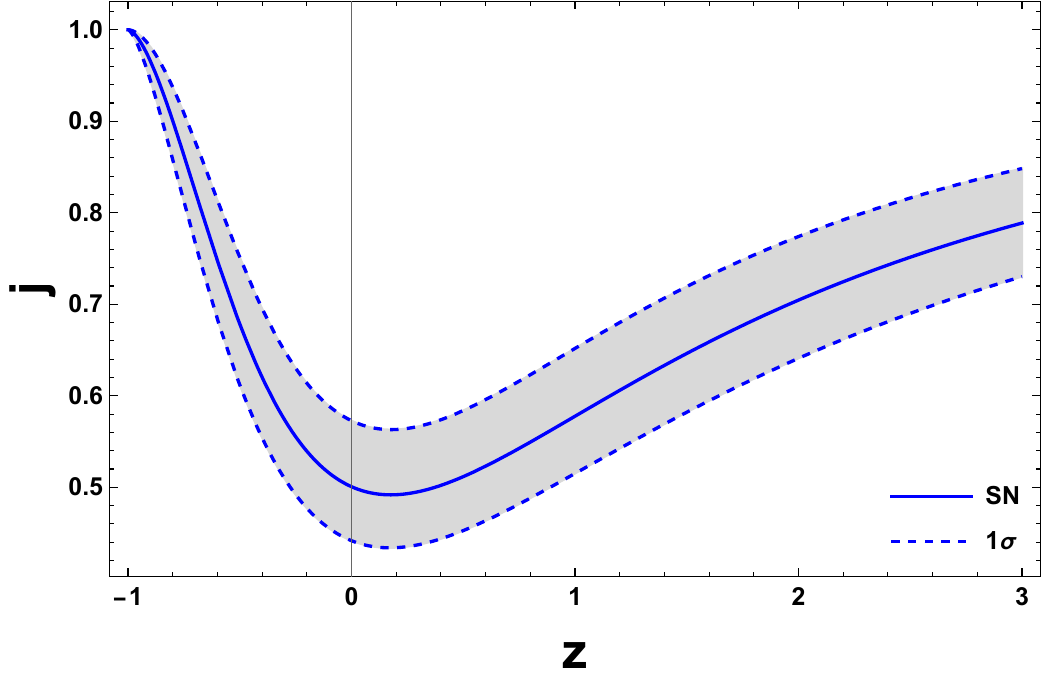}}
\hspace{0.1in}
\subfigure[]{\includegraphics[width=0.3\linewidth]{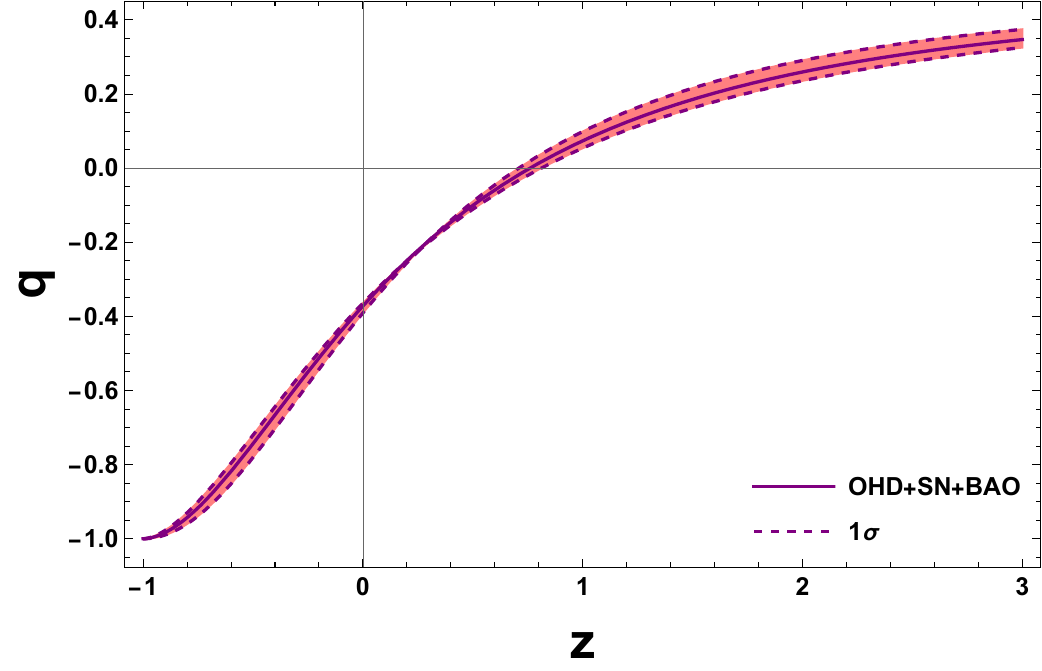}}\hspace{0.1in}
\subfigure[]{\includegraphics[width=0.3\linewidth]{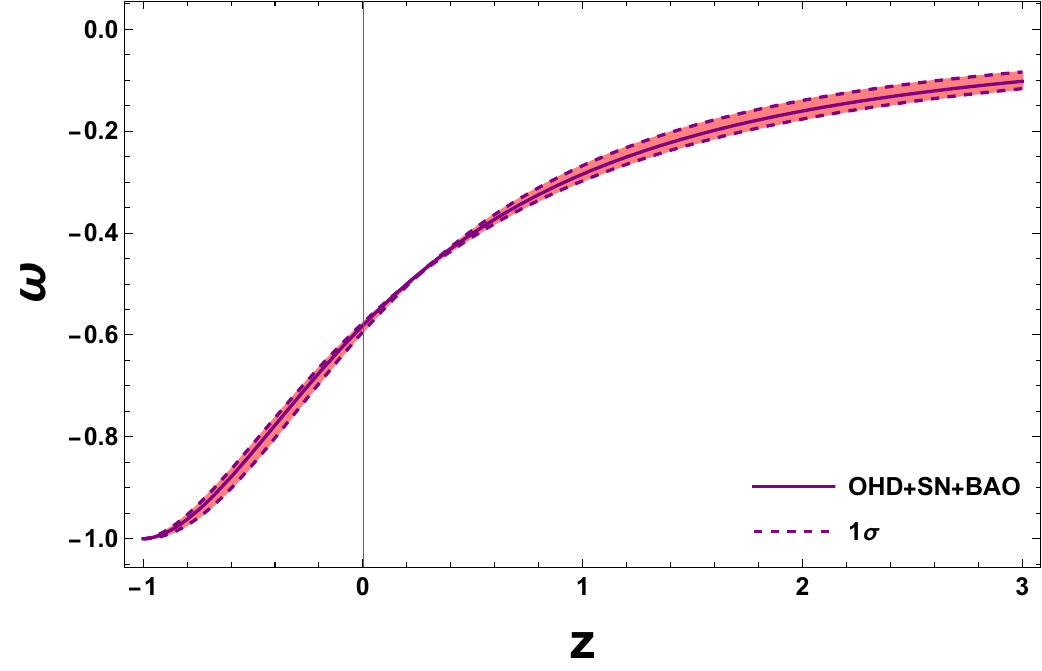}}
\hspace{0.1in}
\subfigure[]{\includegraphics[width=0.3\linewidth]{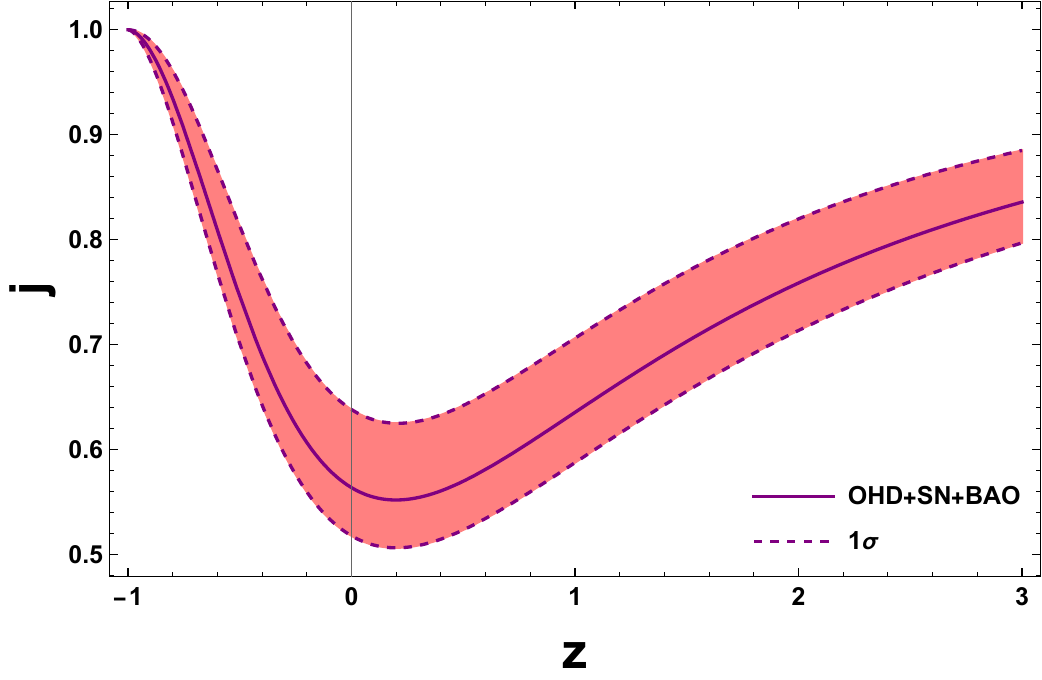}}
\caption{Behavior of the deceleration parameter, the equation of state parameter and the jerk parameter for the case of $q_{f}=-1$. The upper panel shows the behavior for $SN$, and the lower panel is for $OHD+SN+BAO$.}
\label{fig4}
\end{figure}

\end{widetext}

Let us consider the case in which, besides fixing $q_{i}=1/2$, the final value of the deceleration parameter is fixed to the value $q_{f}=-1$, such that asymptotically in the future, i.e., $z=-1$, the model approaches to a de-sitter phase. Models such as $\Lambda$CDM, DGP, quintessence Chaplygin gas, etc., belong to this class of models \cite{Dvali}. By fixing $q_{f}$, we are left with only three free parameters, namely $(H_{0},z_{t},\alpha)$. We present the $68\%$ and $95\%$ confidence levels for this parameter space in FIG. \ref{fig3}. The present values of $q$ for $SN$ and $OHD+SN+BAO$ are obtained explicitly using equation \eqref{qf}, as mentioned in Table \ref{table1}. \\
We plot the deceleration parameter $q(z)$, equation of state parameter $\omega(z)$, and jerk parameter $j(z)$ for all these considered data in FIG. \ref{fig4}. The behavior of $q(z)$ shows a signature flip and $z_t$ indicates the epoch when the expansion of the universe went from a decelerating to an accelerating phase and approaches a de-sitter phase as expected. The present values of $\omega$ are obtained as $-0.56^{+0.007}_{-0.008}$, $-0.58^{+0.005}_{-0.012}$ for $SN$ and $OHD+SN+BAO$, respectively. It reveals that the $\omega$ remains greater than $-1$ (quintessence phase). It does not cross a phantom divide for the far future in this case. 

For this model, the jerk parameter $j(z)$ is found to be evolving, which indicates a tendency of deviation of the universe from the standard $\Lambda$CDM model. Observations indicate that, in this model, the current value of $j$ is less than $1$ for all the datasets ($j_{0}=0.5$ and $0.56$ for $SN$, $OHD+SN+BAO$, respectively) \cite{Mamon/2018}. Therefore, based on the existing model (where $j \neq 1$ ), it is evident that the present model seems to converge to $1$, i.e.,  the standard $\Lambda$CDM in the near future. This would indicate that the higher value of $H_{0}$ leads to a decreased jerk constraint and increased $z_{t}$. The significant deviations from the $\Lambda$CDM predictions suggest a preference for using a parametrization of the dynamical parameter for understanding the dark sector of the universe. It should be noted that the present values of $q_{0}$, $H_{0}$, $z_{t}$ are consistent with the values obtained by many authors \cite{Capozziello/2022,Mukherjee,H0} (also shown in FIGs. \ref{fig8}). 

\subsection{When $\alpha=1/3$}

\begin{figure}[]
\centering
\includegraphics[scale=0.65]{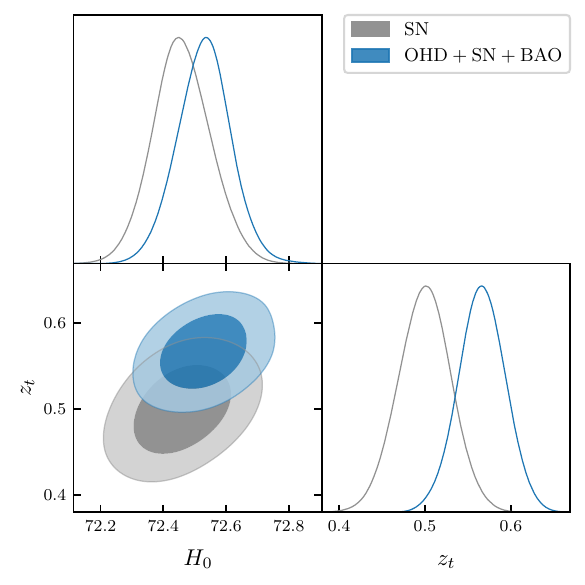}
\caption{The marginalized posterior distributions in the parameter space $(H_{0}, z_{t})$ for the case when $\alpha = 1/3$.}
\label{fig5}
\end{figure}

\begin{figure}[H]
\centering
\includegraphics[scale=0.45]{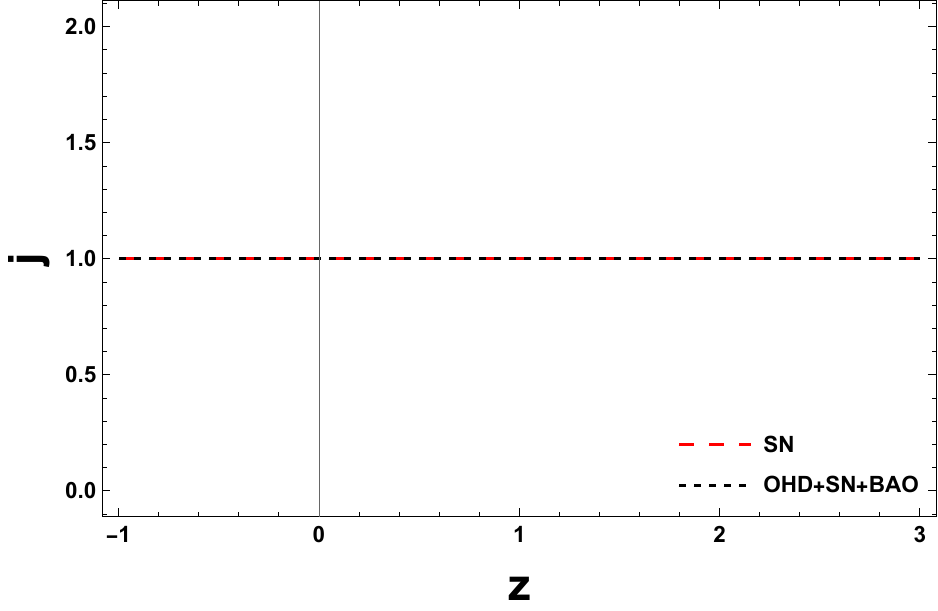}
\caption{Behavior of the jerk parameter when $\alpha = 1/3$ for $SN$ and $OHD+SN+BAO$.}
\label{fig7}
\end{figure}

We also considered the case of $\Lambda$CDM, when $\alpha$ is also fixed, i.e. $\alpha=1/3$. The parameter space here reduces to $(H_{0}, z_{t})$. We have obtained $q_{0}$ from the same equation \eqref{qf}, mentioned in Table \ref{table1}. We present the $68\%$ and $95\%$ confidence levels for this parameter space in FIG. \ref{fig5}. It is clear that the $\Lambda$CDM is in good agreement with the data and used for comparing with the two previous cases. Here, the measurements are giving us tighter constraints on the parameters. The behavior of $q(z)$ shows a signature flip and $z_t$ indicates the epoch when the expansion of the universe went from a decelerating to an accelerating phase in the recent past and approaches a de-sitter phase as expected. In this case, the jerk parameter is $j=1$ for the best-fit values, consistent with the observations (see FIG. \ref{fig7}). 

In FIG. \ref{fig6}, we have shown the evolution of the Hubble parameter $H(z)$ as a function of redshift $z$ compared with 31 $H(z)$ measurements. We have plotted the data points for $H(z)$ with $1\sigma$ error bars. It is observed that our model is well consistent with the $H(z)$ data against redshift for free $q_{f}$, $q_{f}=-1$, $\alpha=1/3$. Furthermore, the behavior of $\mu(z)$ (for Pantheon+) is seen in FIG. \ref{fig6} for all three cases.

\begin{widetext}

\begin{table}[H]
\begin{center}
\label{table1}
{
\begin{tabular}{l c c c c c c c c}
Model & Datasets& $H_0$ & $q_{0}$ & $z_{t}$ & $\alpha$ & $\chi_{\rm min}^2$ & $\rm AIC$ & $\rm BIC$ \\
\hline
$q_{f}$ is free & SN & $72.126\pm 0.096$ & $-0.298 ^{+0.032}_{-0.028}$   & $0.811^{+0.093}_{-0.082}$ & $0.236^{+0.085}_{-0.097}$ & $1605.06$ & $1613.06$  & $1634.81$ \\ 
& Joint  & $72.137\pm 0.098$ & $-0.313\pm 0.02$   & $0.823 \pm 0.065$ & $0.205 \pm 0.067$ & $1633.133$ & $1641.133$ & $1662.97$ \\ 
\hline

Model & Datasets& $H_0$ & $q_{0}$ & $z_{t}$ & $\alpha$ & $\chi_{\rm min}^2$ & $\rm AIC$ & $\rm BIC$ \\
\hline
$q_{f}=-1$ & SN   & $72.250\pm0.096$ & $-0.349^{+0.011}_{-0.013}$   & $0.792\pm 0.073$  & $0.608 \pm 0.063$ & $1605.328$ & $1611.328$ & $1627.644$ \\ 
& Joint  & $72.199\pm0.092$ & $-0.371^{+0.007}_{-0.018}$   & $0.760 \pm 0.051$ & $0.554^{+0.040}_{-0.054}$ & $1634.995$ & $1640.995$ & $1657.37$  \\ 
\hline

Model & Datasets& $H_0$ & $q_{0}$ & $z_{t}$ & $\alpha$ & $\chi_{\rm min}^2$ & $\rm AIC$ & $\rm BIC$ \\
\hline
$\alpha$ fixed & SN   & $72.459\pm0.092$ & $-0.441^{+0.022}_{-0.021}$   & $0.500\pm 0.031$ & $1/3$ & $1609.89$ & $1613.89$ & $1624.76$ \\ 
& Joint  & $72.530\pm0.086$ & $-0.486 \pm 0.017$ & $0.566\pm 0.027$ & $1/3$ & $1642.834$ & $1646.834$ & $1657.75$ \\ 
\hline

\end{tabular}
}
\caption{Summary of the best-fit values of model parameters and statistical analysis using SN and OHD+SN+BAO, including the confidence levels.}
\label{table1}
\end{center}
\end{table}

\begin{figure}[]
\centering
\subfigure[]{\includegraphics[width=0.45\linewidth]{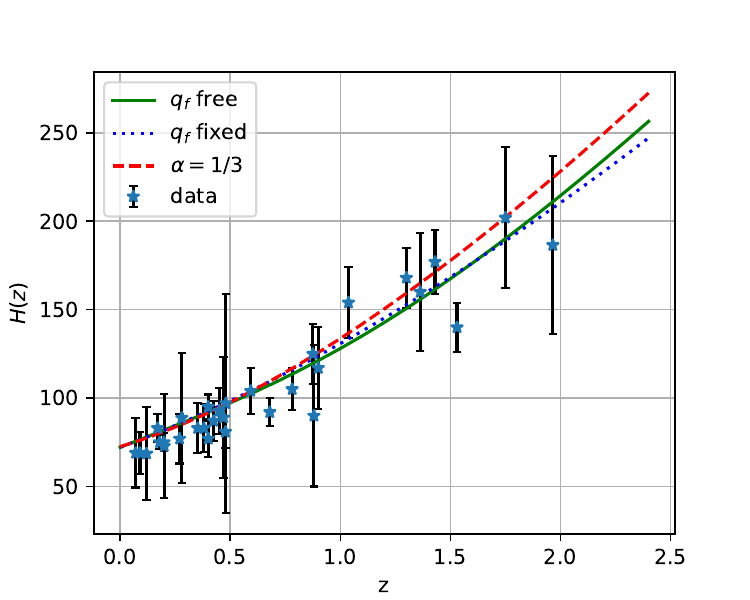}}
\hspace{0.1in}
\subfigure[]{\includegraphics[width=0.45\linewidth]{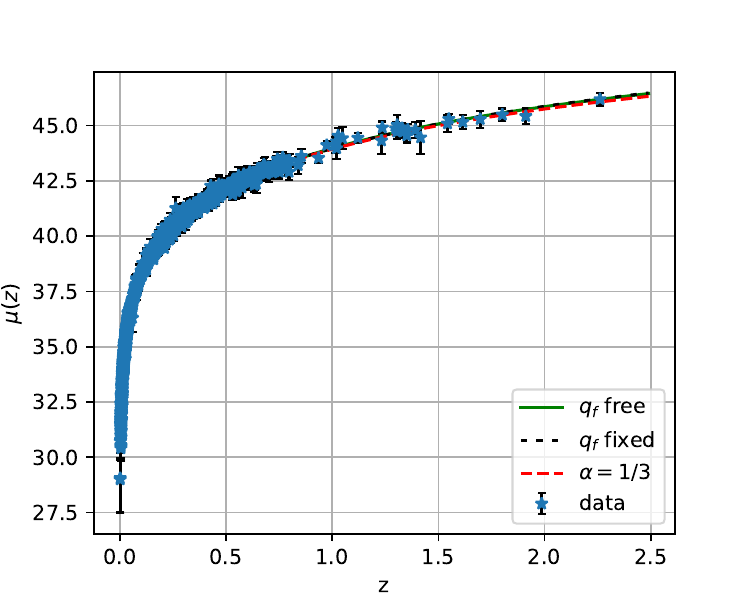}}
\caption{The evolution of Hubble parameter $H(z)$ and $\mu(z)$ as a function of $z$. We have plotted $H(z)$ measurements and Pantheon+ measurements with its error bars for all the cases.}
\label{fig6}
\end{figure}

\end{widetext}

\begin{widetext}

\begin{figure}[]
\centering
\subfigure[]{\includegraphics[width=0.3\linewidth]{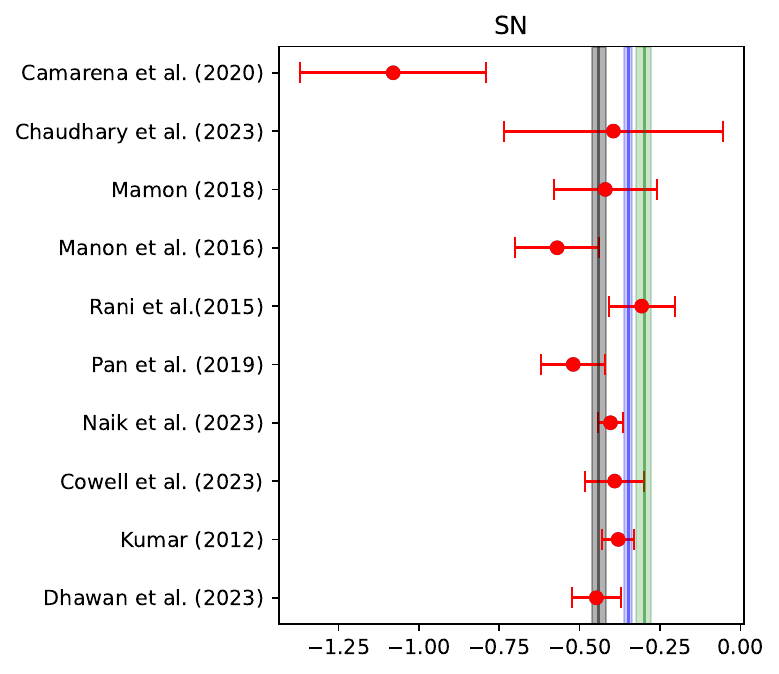}} \hspace{0.1in}
\subfigure[]{\includegraphics[width=0.3\linewidth]{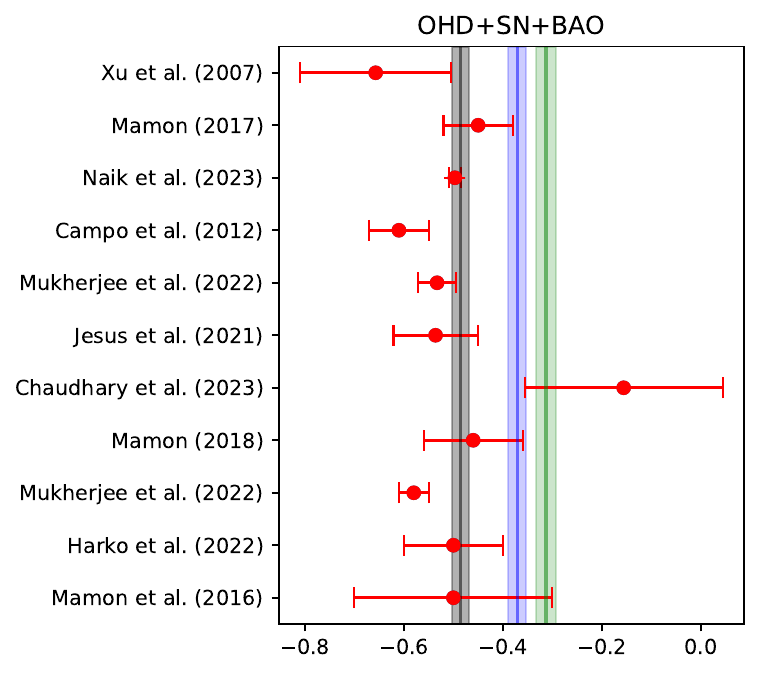}}
\hspace{0.1in}
\includegraphics[width=0.3\linewidth]{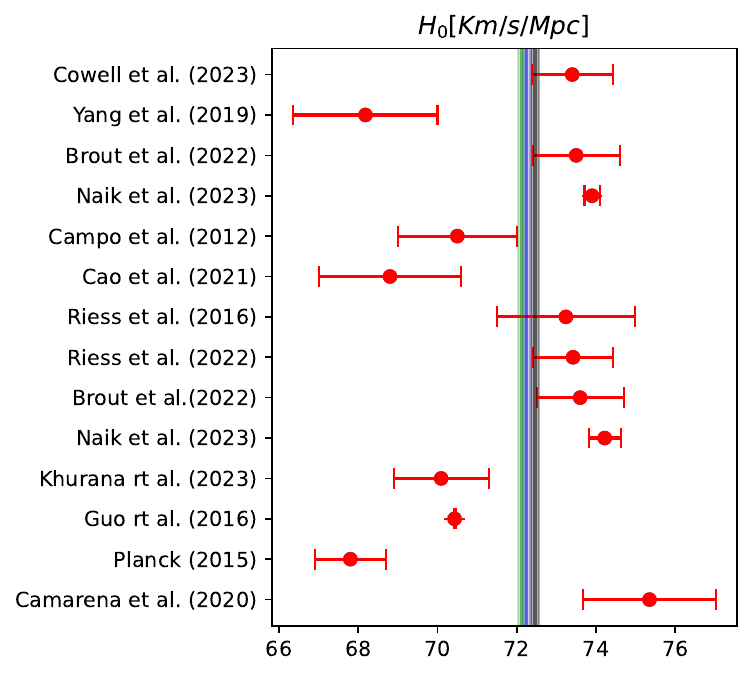}
\caption{The plots represent the present values of $q_{0}$ and $H_{0}$ with their corresponding error bars obtained from several works. The green, blue, and black lines represent the best-fit estimated values obtained for the current model in all three scenarios (free $q_{f}$, $q_{f}=-1$, $\alpha=1/3$), with the related $1\sigma$ error shown as shaded portions \cite{Mukherjee,H0}.}
\label{fig8}
\end{figure}

\end{widetext}

\section{Conclusion}
\label{section 5}

In this work, we reexamined a kink-like parametrization for the deceleration parameter to investigate the transition from cosmic deceleration to acceleration in a spatially flat FLRW universe. The functional form of $q(z)$ involves $q_f$, $q_i$, $q_0$, $z_t$, $\alpha$, and $H_0$ as free parameters. Given that our parametrization specifically intends to explain the cosmic evolution that begins with a matter-dominated decelerated phase, we set the value of $q_i=1/2$. However, as new data pour in and new techniques evolve, revisiting the nature
of $q$ with newer datasets is imperative. So, here, we focussed on recent OHD, SN (Pantheon+ samples), and BAO observations and carried out an MCMC analysis of the model, which allowed us to determine the optimal values of the model parameters. \\
We first consider the case in which we do not impose a fixed value for $q_{f}$. So, we constrained four parameters namely $(H_{0},q_{0},z_{t},\alpha)$. The comparison of the observational data on the $SN$ shows a good agreement with the $\Lambda$CDM model. The AIC analysis also substantiates the presence of a little favor for $OHD+SN+BAO$ between the current model for this scenario and the $\Lambda$CDM predictions. However, to definitively address this matter, additional observational evidence covering a more comprehensive range of redshifts is required. There is an uninspired phase of the universe when $z \rightarrow -1$ since $q(z)$ diverges more from the de-sitter phase for these datasets. Another choice for $q_f$ could also be made to ensure the model's consistency in the distant future. The evolution of the equation of state parameter reveals that the $\omega$ is positive at higher redshifts and remains greater than $-1$ (quintessence phase) until the present time but crosses a phantom divide for the far future. The present value of $\omega$ is $-0.532^{+0.021}_{-0.018}$ and $-0.54 \pm 0.13$, for $SN$ and $OHD+SN+BAO$, respectively. Observations indicate that, in this model, the current value of $j$ is far from $1$ for all the datasets ($j_{0}=0.05$ and $0.02$ for $SN$, $OHD+SN+BAO$, respectively). But in this case, the $j(z)$ is not well constrained as it may be connected to the appearance of abrupt future singularities.\\
We also examined the scenario where the model exhibits the final de-Sitter phase, characterized by $q_{f}=-1$. The Pantheon+ samples are the primary source for improvements compared to prior works. It is anticipated that these samples lead to a substantial enhancement in the constraints, owing to their higher statistics. By incorporating these datasets, we analyze the influence of $H_{0}$ on the transition $z_{t}$. We find that the cosmological deceleration-acceleration transition depends on the value of $H_{0}$ and is also mildly dependent on other cosmological parameters. Here, the jerk parameter is well constrained as it converges to $j_{0}=1$ at late times.\\
We also considered the case of $\Lambda$CDM, when $\alpha$ is also fixed, i.e. $\alpha=1/3$. The parameter space here reduces to $(H_{0}, z_{t})$.
The values of cosmological parameters, including the jerk parameter $j=1$, are consistent with the observations considered. \\
Our results have shown that the present model may represent an interesting alternative to dark energy for the case of $q_{f}=-1$ to tighten the constraints on the parameters for $OHD$, $Pantheon+$, and $BAO$. Our current research reveals a deviation of our model from the standard $\Lambda$CDM when considering a combined dataset, indicating a preference for utilizing a parametrization of the dynamical parameter. As a result, there is an opportunity to delve deeper into understanding cosmological tensions and their resolutions, necessitating the development of new cosmological probes beyond our current study. In the future, we aim to incorporate new observational datasets such as BINGO, LSST, and DESI to investigate cosmological tensions further.

\section*{Data Availability Statement}
There are no new data associated with this article. 

\section*{Acknowledgments}

SA \& PKS thank IUCAA, Pune, India, for the hospitality where part of the work was carried out. PKS  acknowledges the Science and Engineering Research Board, Department of Science and Technology, Government of India, for financial support to carry out Research Project No.: CRG/2022/001847. We are very much grateful to the honorable referee and to the editor for the illuminating suggestions that have significantly improved our work in terms of research quality and presentation.

\end{document}